# Metal-insulator transition via control of spin liquidity in a doped Mott insulator

Hiroshi Oike[1*], Kazuya Miyagawa[1], Hiromi Taniguchi[2], Hiroyuki Okamoto[3], Kazushi Kanoda[1*]

[1] *Department of Applied Physics, University of Tokyo, Bunkyo-ku, Tokyo 113-0032, Japan.*

[2] *Department of Physics, Saitama University, Saitama, Saitama 338-8570, Japan.*

[3] *Department of Quantum Medical Technology, Kanazawa University, Kanazawa 920-0942, Japan*

*Corresponding author. E-mail: oike@ap.t.u-tokyo.ac.jp (H.O.); kanoda@ap.t.u-tokyo.ac.jp (K.K.)

**Quantum spin liquid states, in which spins are quantum-mechanically delocalized in direction, have been so far studied for charge-localized Mott insulators arising from strong repulsive interaction[1-11]. Recently, however, it was found that the doped Mott insulator with a triangular lattice, $\kappa$-(ET)$_4$Hg$_{2.89}$Br$_8$, exhibits both spin-liquid-like magnetism[12] and metallic electrical conduction[12-17]. Thus, it is now possible to experimentally explore how the spin liquidity affects the electrical conduction, an issue that has received a great deal of theoretical attention[9-11,18-20]. Here, with a newly developed method to combine uniaxial and hydrostatic pressures, we investigate the electrical conduction in the doped Mott insulator with controlling the triangular lattice geometry and the repulsion strength which determines the spin-liquidity[1-8] and Mottness[21-23], respectively. We found that, in a strongly interacting regime, the electronic state drastically changes from an insulator to a Fermi liquid via a non-Fermi liquid with varying geometrical frustration, which suggests that spin liquidity promotes delocalization of charges. This result indicates that frustration in spin degrees of freedom has a decisive impact on the transport of charges through the entanglement of spin and charge in a doped Mott insulator.**

Correlated electrons in solids harbor a range of emergent phenomena stemming from the competition between localization and delocalization. The particle-like charge localization is induced by mutual interactions as best manifested in the Mott insulator or Wigner crystal, however, which is delocalized into extended waves by reducing the interactions, or partially removing or adding electrons[24]. Diverse phenomena of current interest such as unconventional superconductivity emerge on the verge between localization and delocalization, regulated by the interaction strength or carrier doping[12-26]. On the other hand, spin degrees of freedom also exhibit a similar competition in a different way. Akin to the localization of charges, interacting spins usually fix their directions in a classical manner at low temperatures, where quantum nature barely appears as spin contraction[27]. However, when spins stay on lattices of particular geometries like triangular lattices where inter-site interactions are geometrically frustrated, the quantum fluctuations are vitalized and most highlighted by the emergence of the quantum spin liquid (QSL) free from the classical orders even at absolute zero[1-11]. Thus, interaction strength, doping and frustration are key parameters to emergent phenomena, which have been indeed fertilized through varying these parameters albeit independently so far. If they are controlled in a combined manner, it would open new parameter space for exploring electronic phases and phenomena because, in strongly correlated electron systems, charge and spin degrees of freedom are intertwined[28-32]. As an experimental challenge in this line, the present work investigates a doped QSL candidate under the combined variation of interaction and frustration, using a sophisticated pressure technique developed here. A global metal-insulator phase diagram illustrating highly susceptible charge-state in the doped QSL to frustration as well as interaction is revealed.

The material investigated here is the layered organic conductor, $\kappa$-(ET)$_4$Hg$_{2.89}$Br$_8$ ($\kappa$-HgBr hereafter), comprised of conducting ET layers and insulating Hg$_{2.89}$Br$_8$ layers (Fig. 1**a**)[13,33,34]. In the ET layer, ET dimers form a triangular lattice represented by the transfer integrals of $t$ and $t'$ with $t'/t$ of 1.02, suggesting a nearly isotropic triangular lattice (Fig. 1**b**, **c**)[7,26]. The ratio of the on-dimer-site Coulomb repulsion $U$ to the bandwidth $W$ is 1.11, which exceeds the value for the triangular-lattice QSL Mott insulator, $\kappa$-(ET)$_2$Cu$_2$(CN)$_3$ ($\kappa$-Cu$_2$(CN)$_3$ hereafter)[1,12,15]. Noticeably, the nonstoichiometric Hg$^{2+}$ ions forming a fixed incommensurate lattice with the ET dimer lattice shifts the filling of the antibonding band of the ET dimers from half-filling by 11%[33,34], as confirmed by Raman spectroscopy[35]. According to previous studies, $\kappa$-HgBr is a non-Fermi liquid metal with spin susceptibility nearly perfectly scaled to that of the QSL candidate, $\kappa$-Cu$_2$(CN)$_3$, and thus is a doped QSL candidate[12]. From experimental point of view, the high compressibility of organic materials is an advantage for the pressure control of electronic states[22,23]. Previous experiments revealed that hydrostatic pressure induces a non-Fermi liquid to Fermi liquid crossover in the normal state[15,16] and a BEC to BCS crossover in the superconducting condensate[16]. The application of uniaxial pressure deforms the triangular lattice and is expected to vary frustration strength[1-8, 36].

We simultaneously compress and deform the triangular lattice of $\kappa$-HgBr by hydrostatic and uniaxial pressures combined in a controllable manner. A single crystal of $\kappa$-HgBr is placed in a conventional piston cylinder cell filled with Demnum oil s-20, a liquid pressure medium, which solidifies on cooling (for more details, see methods). Thus, pressure applied by pressing the piston is hydrostatic when the oil remains liquid (blue region in Fig.1**d**) but is uniaxial under the solidified oil at high pressure and/or low temperature (yellow region in Fig. 1**d**). In pressure sequences #1 and #2, we push the piston at room temperature until the piston pressure reaches a target value $P_t$ exceeding 1.1 GPa, at which the resistance of the strain gauges confirms that the oil has solidified (Supplementary Fig. 1**a**); in these processes, the sample is under hydrostatic pressure up to 1.1 GPa and then subjected to a uniaxial pressure of $P_t - 1.1$ GPa. In pressure-temperature sequences #3 and #4, we initially push the piston at ambient temperature until a target hydrostatic pressure (< 1.1 GPa) was reached. Then, the pressure cell is cooled down to a temperature below the solidification point of the oil and further pushed the piston up to a target piston pressure $P_t$ (for more details, see methods); in the sequence #3 (#4), a hydrostatic pressure of 0.5 GPa (0.2 GPa) and a uniaxial pressure of $P_t - 0.5$ GPa ($P_t - 0.2$ GPa) were applied (Supplementary Fig. 1**b**). This pressurization method allows one to simultaneously apply hydrostatic pressure $P_h$ and uniaxial pressure $P_u$ in a combined way (Fig. 1**e**).

The application of hydrostatic pressure to $\kappa$-HgBr inevitably causes deformation as well as compression of the triangular lattice and likewise the uniaxial pressure induces compression as well. Thus, it is required to relate the pressure values ($P_h$, $P_u$) to the parameters ($U/W$, $t'/t$) based on the structure and band calculation. Because of difficulty in experimentally determining the exact atomic configurations in the present pressure conditions, we made the following assumptions on the molecular arrangements under pressure. Firstly, we determined the lattice constants of $\kappa$-HgBr with using the reported compressibility values of $\kappa$-(ET)$_2$Cu(NCS)$_2$ under hydrostatic pressure and the strain values separately measured with strain gauge under uniaxial pressure. Then, we determined the molecular arrangements, assuming that ET molecules are so rigid as not to deform, the relative coordinates of the molecular centers are invariant and the orientation of ET changes by the same angle as the arctan($b/c$), where $b$ and $c$ are the lattice constants of the b- and c-axes. Using the molecular arrangements thus determined, we calculated the transfer integrals between the molecular orbitals by the extended Hückel plus tight binding method, which gave the values of $U$, $W$, $t$, and $t'$ of the dimer model. Thus, we could locate the pressure sequences #1-4 in the $U/W$-$t'/t$ plane as shown in Fig. 1**f**. The detail and validity of the present method is discussed in supplementary information.

We compare the temperature variation of in-plane resistivity $\rho_{//}$ under different pressure conditions, ($P_\mathrm{h}$, $P_\mathrm{u}$), prepared by the pressurization sequences for generating uniaxial strains parallel to b-axis (#1) and c-axis (#2-4) (Fig. 1**f**). All the samples used in these sequences reproduced the previous results under hydrostatic pressure. The behaviour of $\rho_{//}$ strongly depends on the direction of the applied uniaxial strain as well as on the value of $P_\mathrm{h}$ indicating that the electronic phase depends on both uniaxial and hydrostatic pressures.

First, we examine the directional dependence of the uniaxial strain effect on $\rho_{//}$. Figure 2**a** shows the evolution of the temperature variation of $\rho_{//}$ after the pressure sequences #1 and #2, where pressure up to 1.1 GPa is hydrostatic and further pressure exceeding 1.1 GPa gives uniaxial strains parallel to b-axis (#1) and c-axis (#2) with $P_\mathrm{h}$ = 1.1 GPa. At hydrostatic pressures up to 1.1 GPa, $\rho_{//}$ shows metallic behavior with smaller values at higher $P_\mathrm{h}$. Under b-axis strain caused by further press ($P_\mathrm{u}^\mathrm{b}$ = 0.3, 0.6, 0.9 and 1.3 GPa with $P_\mathrm{h}$ = 1.1 GPa on sequence #1), the metallic behavior is further enhanced with the increase of $P_\mathrm{u}$. However, under c-axis strain ($P_\mathrm{u}^\mathrm{c}$ = 0.2, 0.5, 0.8, 1.1 and 1.3 GPa with $P_\mathrm{h}$ = 1.1 GPa on sequence #2), non-metallic behaviour appears at high temperatures in a high-$P_\mathrm{u}$ range, (Fig.

2**a**). To characterize the temperature dependence of $\rho_{//}$ in the low-temperature metallic state, we examined the temperature-dependent exponent $\alpha$ by fitting the form of $\rho_{//} = \rho_o + AT^\alpha$ to the experimental data for various $P_u$ values with $P_h = 1.1$ GPa below 15 K. As seen in Fig. 2**b**, the c-axis pressurization makes the system trend toward a non-Fermi liquid ($\alpha = 1$), whereas the b-axis pressurization stabilizes a Fermi liquid ($\alpha = 2$). Evidently, electrons show contrasting responses to the distortion direction of the triangular lattice.

The appearance of non-metallic behavior under c-axis strain was totally unexpected. Thus, we further examined the c-axis strain effect with lower values of $P_h = 0.5$ GPa ($P_u^c = 0.5, 0.7, 1.0, 1.3, 1.6$ and $2.0$ GPa with $P_h = 0.5$ GPa on sequence #3) and $P_h = 0.2$ GPa ($P_u^c = 0.9, 1.1, 1.4$ and $1.7$ GPa with $P_h = 0.2$ GPa on sequence #4); namely, the triangular lattice is distorted under more strongly interacting conditions. The results are shown in Fig.3: note that the scale of the vertical axis is logarithmic. As $P_h$ is lowered, the magnitude of $\rho_{//}$ increases and exceeds 1 $\Omega$ cm under uniaxial pressure with $P_h$ of 0.2 GPa (Fig. 3). The $\rho_{//}$ values at room temperature are on the order of $10^{-2}$ $\Omega$ cm in all pressures measured; however, $\rho_{//}$ increases by two orders of magnitude on cooling from 300 K down to 100 K, indicating the emergence of an insulating state with a charge gap of about 1000 K (Supplementary Fig. 2), which is comparable to the values of the Mott insulating phases in non-doped $\kappa$-ET compounds. The resistivity drops at low temperatures possibly come from a tiny fraction of metallic domains; the residual resistivity of ~1 $\Omega$ cm under $P_u^c = 0.9, 1.1, 1.4$ and $1.7$ GPa with $P_h = 0.2$ GP is two or three orders of magnitude over the Mott-Ioffe-Regel limit, ruling out the bulk metallicity.

To situate the low-temperature electronic states in the $U/W$-$t'/t$ plane, we made a contour plot of the $\rho_{//}$ values at 10 K in this plane (Fig. 4**a**). $\rho_{//}$ takes large values in a low t'/t region and decreases with increasing $t'/t$, indicating the pronounced frustration-dependence of the ground state of $\kappa$-HgBr. Figure 4**b** represents the $t'/t$ dependences of the $\rho_{//}$ values at $U/W \sim 1$ in the contour plot and the temperature exponent $\alpha$ of $\rho_{//}$ and the inset shows the temperature dependence of $\rho_{//}$ for $t'/t$ =0.87, 0.89, 0.97 and 1.12 at a fixed $U/W$ value of ~ 1. It is evident that the insulating state emerging at low $t'/t$ crosses over to a non-Fermi liquid and then to a Fermi liquid as $t'/t$ increases; namely, a non-Fermi liquid on the triangular lattice transitions to an insulator when the lattice is distorted towards a square lattice, but to a Fermi liquid when it is oppositely distorted toward the quasi-one-dimensional coupled chains. Thus, the degree of frustration makes enormous influence on the localization/itinerancy nature of charge in the doped Mott insulator.

Finally, we discuss the origin of the emergence of the insulating state in the low-$t'/t$ region, where the spin frustration is released. In $\kappa$-ET compounds with half-filled bands, a spin liquid insulator is predicted to change to an antiferromagnetic insulator as $t'/t$ decreases[2,3,6,7]. Additional charge carriers to the Mott insulators do not significantly cost extra energy for charge transfer in a background with translational symmetry preserved as in a spin liquid (Fig. 4**c**). However, in a antiferromagnetically ordered or correlated background, the travelling of the carriers to the nearest neighbor sites costs several times the spin exchange energy as depicted in Fig. 4**d**. Because the time scale of charge transfer is generally faster than that of magnetic fluctuations, antiferromagnetic fluctuations can cause such a suppression of charge transfer even in the absence of a static antiferromagnetic order. Then, the doped carriers lose mobility and are likely localized through polaron formation[28] or stripe/charge ordering[29] when spin frustration is weakened by the deformation of the triangular lattice.

Since the discovery of high temperature superconductors, carrier doping to Mott insulators has provided a wide variety of electronic states, which arise from the complex quantum mechanical entanglement exhibited by the charge and spin degrees of freedom. The present study shows that the spin frustration controlled by lattice geometry play a decisive role in the behavior of charge carriers. Further studies with controlled lattice deformation would unravel further complex entanglement between charge and spin, leading to the discovery of novel phases of quantum many-body systems.

**Methods**

**Sample preparation and transport measurements under pressure**

A single crystal of $\kappa$-(ET)$_4$Hg$_{2.89}$Br$_8$ ($\kappa$-HgBr), which is grown by the conventional electrochemical methods, is placed in a dual structured clamp-type cell formed by BeCu and NiCrAl cylinders with Demnum oil s-20 as the pressure medium. The relationship between the shape of the sample and the crystal axis was investigated by either X-ray diffraction or angle dependence of electron spin resonance. To evaluate the quality of applied pressure, we put strain gauges (KFLB-2-120-C1-11, Kyowa Electronic Instruments) in the pressure medium together with the sample. In pressure sequences #1 and #2, the pressure is applied at room temperature exceeding the solidification pressure of the pressure medium. The solidification pressure at room temperature is confirmed by the pressure dependence of the resistance of the strain gauges (Supplementary Fig. 1**a**). In pressure sequences #3 and #4, the pressure medium is cooled and solidified by pouring refrigerant (ethanol with dry ice, or liquid nitrogen) into the container containing the pressure cell with a pressure load applied. Because the solidification temperature can be seen in the temperature dependence of the in-plane resistivity of $\kappa$-HgBr (Supplementary Fig. 3), the resistivity was monitored to ensure that the sample space was below the solidification temperature, and then the pressure load was increased to apply uniaxial pressure. To measure the in-plane resistivity of $\kappa$-HgBr under pressure, four gold wires are connected to a sample at electrodes made of carbon paste. The electrodes are aligned in the direction of conducting plane and perpendicular to the pressure axis. Electrical current is applied with a source meter (Keithley 2400), and the voltage is measured with a nanovolt meter (Keithley 2182A).

**Acknowledgements** This work was supported in part by JSPS KAKENHI under Grant Nos. 11J09324, 18K13512, 18H05225, 19H01846, 20K14410, 20K20894, 20KK0060, and 21K18144.

**Author Contributions** H.Oike conducted transport experiments and analysed the data. H.T. grew the single crystals used for the study. H.Okamoto analysed X-ray diffraction data. H.Oike and K.K. wrote the paper. All authors discussed the results and commented on the manuscript. K.K. designed the project.

**Author Information** The authors declare no competing financial interests. Readers are welcome to comment on the online version of the paper. Correspondence and requests for materials should be addressed to H.O. (oike@ap.t.u-tokyo.ac.jp) or K.K. (kanoda@ap.t.u-tokyo.ac.jp).

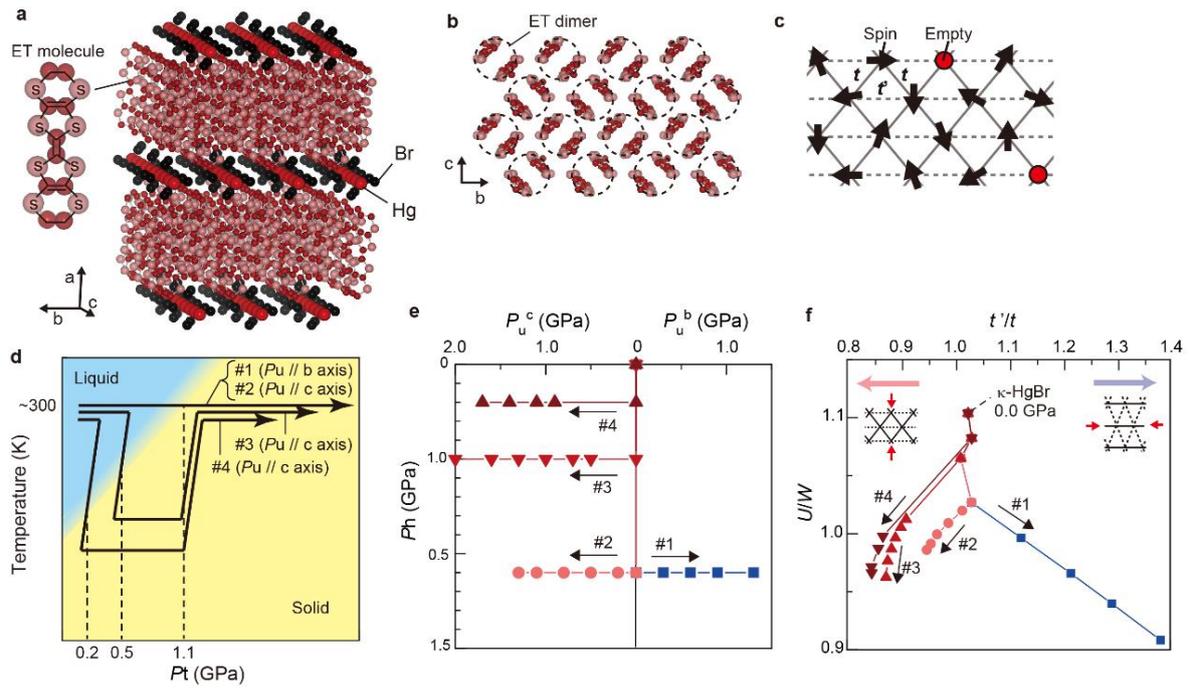

**Figure 1 | Crystal structure and parameter space investigated by the newly developed pressure method. a,** Layered crystal structure of $\kappa$-(ET)$_4$Hg$_{2.89}$Br$_8$ ($\kappa$-HgBr), where ET stands for the molecule, bis(ethylenedithio) tetrathiafulvalene. **b,** Molecular arrangement in the ET layer parallel to the b-c plane. **c,** Schematic representation of spin and charge states in the ET layer. Because the holes are doped to the Mott insulator, where double occupancy is strongly forbidden, the sites occupied by one electron are responsible for the magnetism and the mobile vacant sites contribute to the electrical conduction. **d,** Trajectories of pressure and temperature swept during each pressure sequence. **e,** Values of hydrostatic pressure $P_h$ and uniaxial pressure parallel to b-axis $P_u^b$ (#1) and c-axis $P_u^c$ (#2-4) in four pressure sequences. **f,** Values of $U/W$ and $t'/t$ in the four pressure sequences.

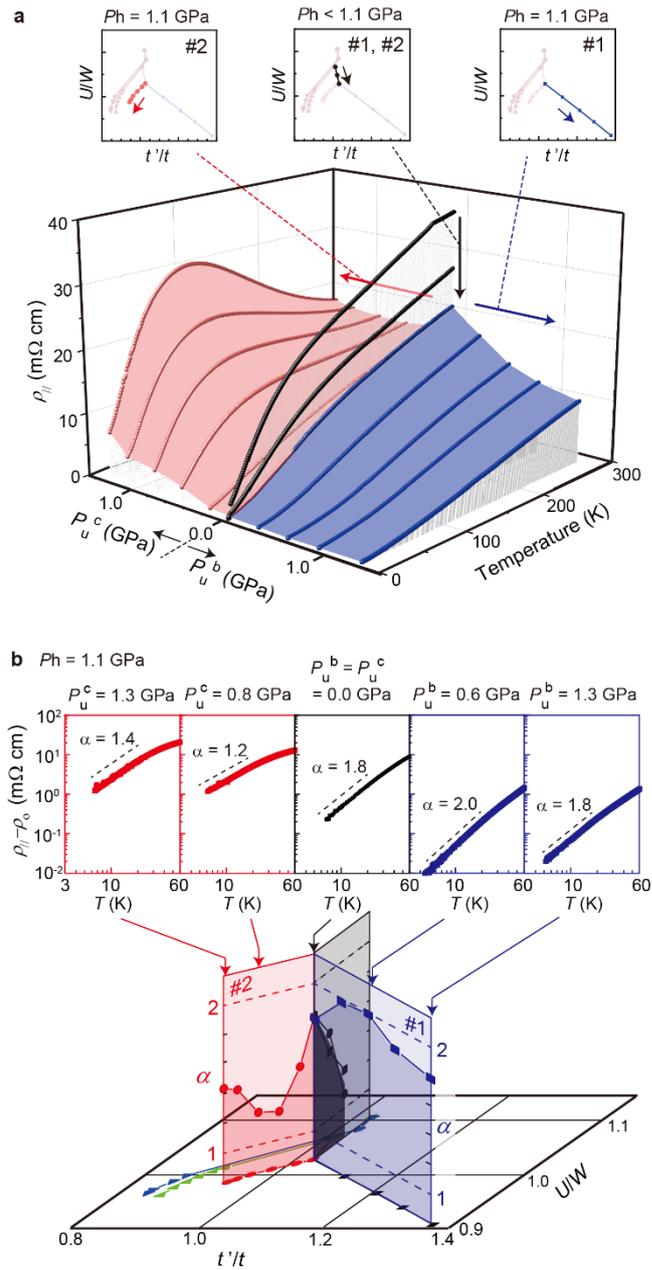

**Figure 2 | Lattice-geometry dependence of metallic nature. a,** Pressure and temperature dependences of in-plane resistivity $\rho_{//}$ in pressure sequences #1 and #2, where pressure up to 1.1 GPa is hydrostatic and further pressure exceeding 1.1 GPa gives uniaxial stress parallel to b-axis $P_u^b$ (#1) and c-axis $P_u^c$ (#2) with $P_h$ = 1.1 GPa. The surface colored in blue (red) corresponds to the pressure sequence #1 (#2). The upper panels show the values of $U/W$ and $t'/t$ for each pressure sequence. **b,**

Pressure dependence of temperature-dependent exponent $\alpha$ in the fit of the form, $\rho_{//} = \rho_0 + AT^{\alpha}$, to the experimental data shown in **a** below 15 K. The upper panels show log ($\rho_{//} - \rho_0$) vs log T at each pressure, which was used to determine the value of $\alpha$.

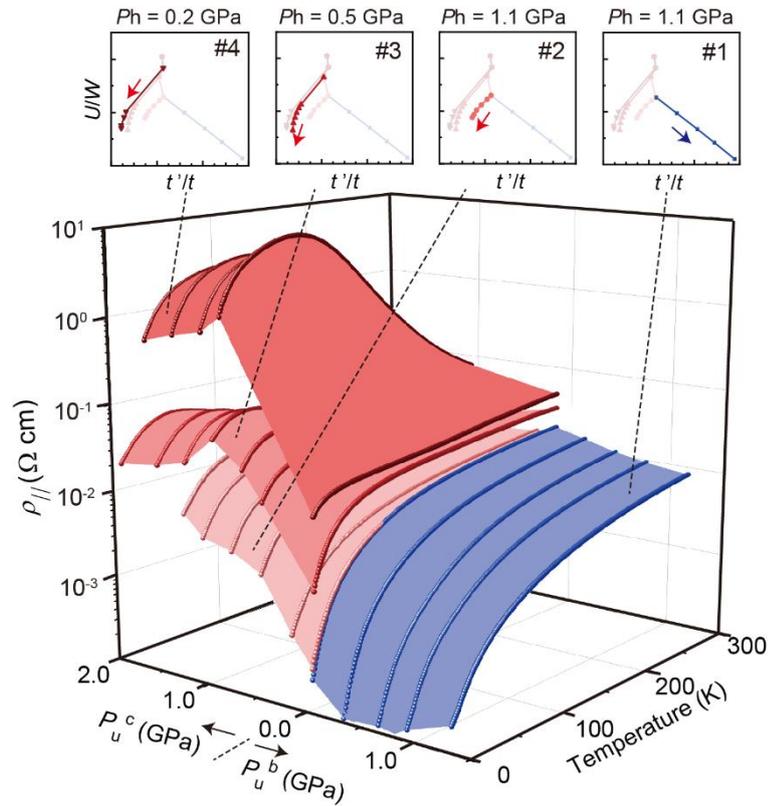

**Figure 3 │ Emergence of insulating phase due to deformation of the triangular lattice.** Pressure and temperature dependences of in-plane resistivity $\rho_{//}$ in four pressure sequences #1-4 plotted in a logarithmic scale. The surface colored in blue corresponds to the pressure sequence #1 and the three surfaces colored in red correspond to the pressure sequence #2-4 with the darker color indicating the smaller $P_h$. The upper four panels show the values of $U/W$ and $t'/t$ for each pressure sequence. In pressure sequence #4, where the triangular lattice is largely deformed, a pronounced increase in $\rho_{//}$ is observed, suggesting the formation of an insulating phase.

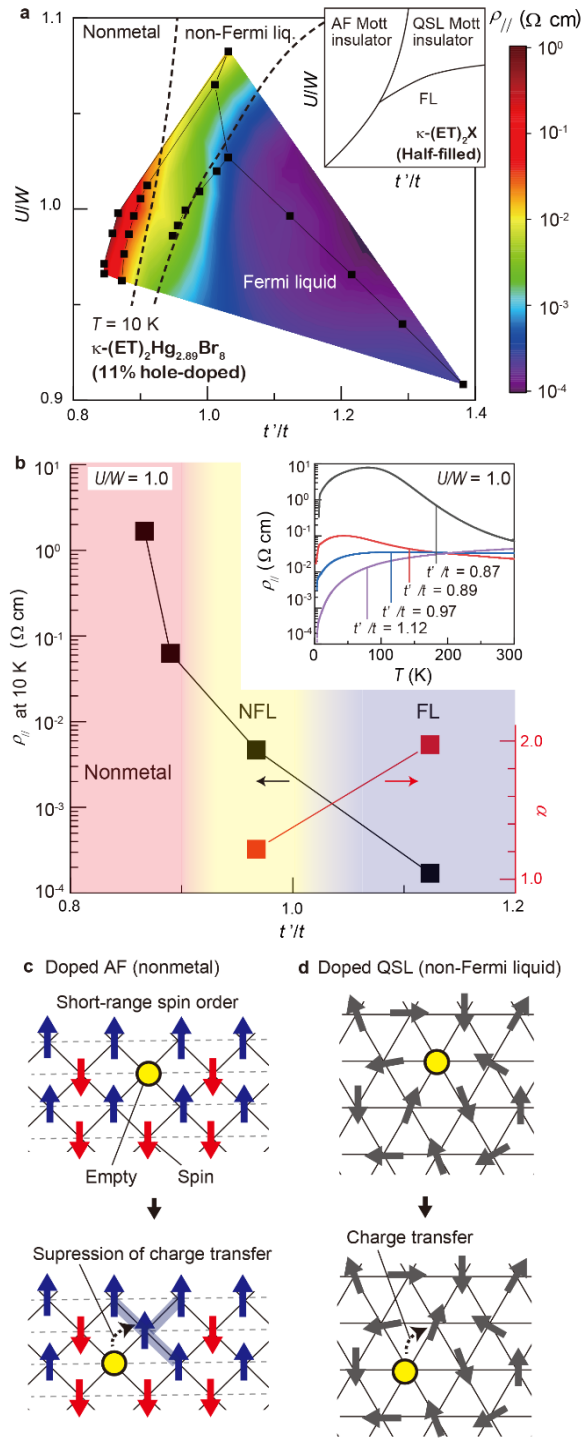

**Figure 4 │Geometrical frustration dependent electrical transport. a,** Contour map of the $\rho_{//}$ values at 10 K on the parameter plane of $U/W$ and $t'/t$. The black squares are the measurement points and the interpolation between them is represented by the colours. The inset shows schematic of theoretically suggested $U/W$-$t'/t$ phase diagram of the half-filled system $\kappa$-(ET)$_2$X. AF, QSL, and FL stand for antiferromagnet, quantum spin liquid, and Fermi liquid, respectively. The non-metal, non-Fermi liquid and Fermi liquid regions in the present doped system correspond to AF, QSL and Fermi liquid regions in the non-doped half-filled case. **b,** $t'/t$ dependence of the $\rho_{//}$ value at 10 K and $\alpha$ value for $U/W$ approximately fixed at 1.0. The inset shows temperature dependence of $\rho_{//}$ for $t'/t$ = 0.87, 0.89, 0.97, and 1.12 with $U/W$ approximately fixed at 1.0. **c, d,** Schematics of doped-carrier transport in the AF (c) and QSL (d) backgrounds. In the presence of AF spin correlation or order, the nearest-neighbor hopping of a carrier accompanies an increase in magnetic energy, which suppresses electrical transport. In the QSL, there occurs no significant change of magnetic energy in quantum disordered spins when a charge carrier hops to adjacent sites.

**Estimation of crystal structure under pressure**

To map the two-dimensional pressure values ($P_h$, $P_u$) to the two-dimensional parameters ($U/W$, $t'/t$), we need information on the molecular arrangement under pressure. Because of difficulty in experimentally determining the exact atomic configurations in the present pressure conditions, assumptions have been made about the molecular arrangements under pressure as described in the main text. These assumptions are explained in more detail below.

Firstly, we determined the lattice constants of $\kappa$-HgBr with using the reported compressibility values of $\kappa$-(ET)$_2$Cu(NCS)$_2$ under hydrostatic pressure [37] and the strain values measured with strain gauges under uniaxial pressure. Figure S4 shows the estimated strain values based on these assumptions. In reality, due to the different crystal structure of the insulating layers, the absolute value of compressibility is not identical to $\kappa$-(ET)$_2$Cu(NCS)$_2$, but this does not affect the qualitative conclusion that the $U/W$ decreases under hydrostatic pressure. In previous studies on ET-based organic conductors under uniaxial pressure, the strain values of the samples are 0.5-0.7 times of those of the strain gauges placed with the samples in a pressure cell [38, 39]. Thus, the amount of strain in the sample may be overestimated. Because the assumptions adopted in the present study include these quantitative imprecisions, we avoid making a conclusion where quantification of the strain value is essential, and discuss the pressure dependences of electronic states in a qualitative manner.

Then, we determined the molecular arrangements, assuming that ET molecules are so rigid as not to deform, the relative coordinates of the molecular centers are invariant and the orientation of ET changes by the same angle as the arctan($b/c$), where $b$ and $c$ are the lattice constants of the b- and c-axes. When we plot $\phi$ dependence of $\theta$ in $\kappa$-(ET)$_2$Cu$_2$(CN)$_3$, $\kappa$-(ET)$_2$Cu[N(CN)$_2$]Cl, $\kappa$-(ET)$_2$Cu [N(CN)$_2$]Br, and $\kappa$-(ET)$_2$Cu[N(CN)$_2$]I [40], these compounds roughly follow the relationship of $\Delta\phi = \Delta\theta$ (Fig. S5**b**), where $\theta$ and $\phi$ are the molecular orientation and arctan($c/b$), respectively, as shown in Fig, S5**a**. The relationship suggests the validity of the assumption that $\Delta\theta$ equals to $\Delta\phi$. Thus, we determined the molecular arrangement under pressure applied with the newly developed method.

**Band calculation**

As we determined all of the atomic positions, we proceeded to the calculation of transfer integrals with the scheme developed by Mori et al. [41], which is based on extend Huckel approximation. First, HOMO (highest occupied molecular orbital) of ET molecule is calculated. Then, the overlap integrals between HOMOs are evaluated. The transfer integral between the HOMO orbitals of ET molecules is estimated by multiplying the overlap integral by $-10$ eV [41]. In $\kappa$-ET compounds, because of the strong dimerization of ET molecules, dimer orbitals form an isosceles-triangular lattice represented by two inter-dimer transfer integrals $t$ and $t'$ [42, 43]. They are given by the equations, $t = (-t_p + t_q)/2$ and $t' = -t_{b2}/2$, $t_p$, $t_q$, and $t_{b2}$ are transfer integrals shown in Fig. S6. $U$ approximately equals to $2t_{b1}$ as described in ref. 42, and $W$ is calculated with the tight-binding approximation, where $U$ is on-dimer Coulomb repulsion and $W$ is bandwidth of conduction band formed by antibonding orbitals of ET dimers. Thus, we obtained the values of $t'/t$ and $U/W$ of interest (Fig. S7).

**Supplementary references**

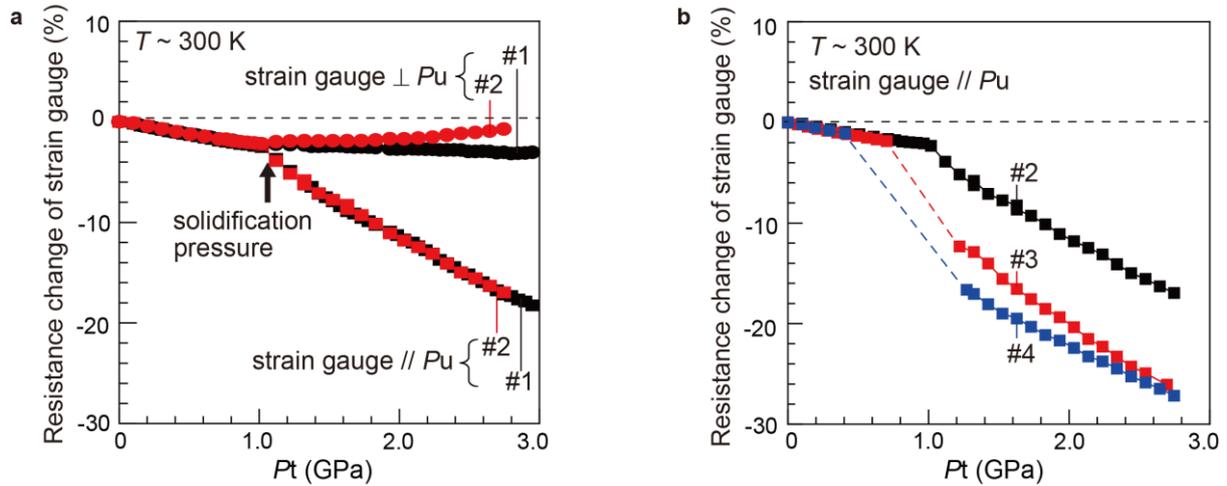

**Supplementary Fig. 1 | Pressure dependence of resistance of strain gauges. a,** Pressure dependences of resistance of strain gauges for pressure sequences #1 and #2, where pressure is applied at room temperature. The strain gauges are set parallel (//$P_u$) or perpendicular (⊥$P_u$) to the pressure axis. The strain gauges set parallel to the pressure axis show almost the same pressure dependences those set perpendicular to it below 1.1 GPa, indicating that the pressure is isotropic. On the other hand, the pressure dependences separate at a pressure above 1.1 GPa, indicating that the pressure is anisotropic. **b,** Pressure dependences of resistance of strain gauges for pressure sequences #2-4, where pressure axis is along $c$ axis. The uniaxial strain at a certain value of $P_t$ gets larger when the values of $P_h$ is lower, indicating that uniaxial component of the applied pressure increases.

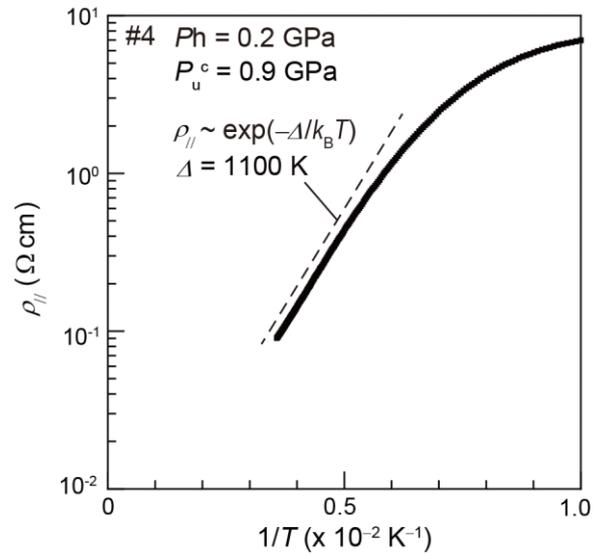

**Supplementary Fig. 2 | Arrhenius plot of temperature dependence of in-plane resistivity $\rho_{//}$.** The temperature dependence between 200 K and 300 K follows the Arrhenius law, and the charge gap in the insulating phase is estimated to be of the order of 1000 K.

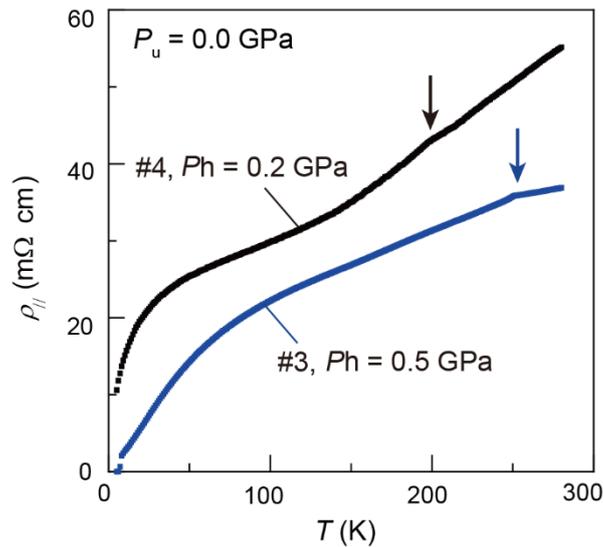

**Supplementary Fig. 3 | Temperature dependence of in-plane resistivity $\rho_{//}$.** The arrows indicate the temperatures below which the pressure medium is in a solid state.

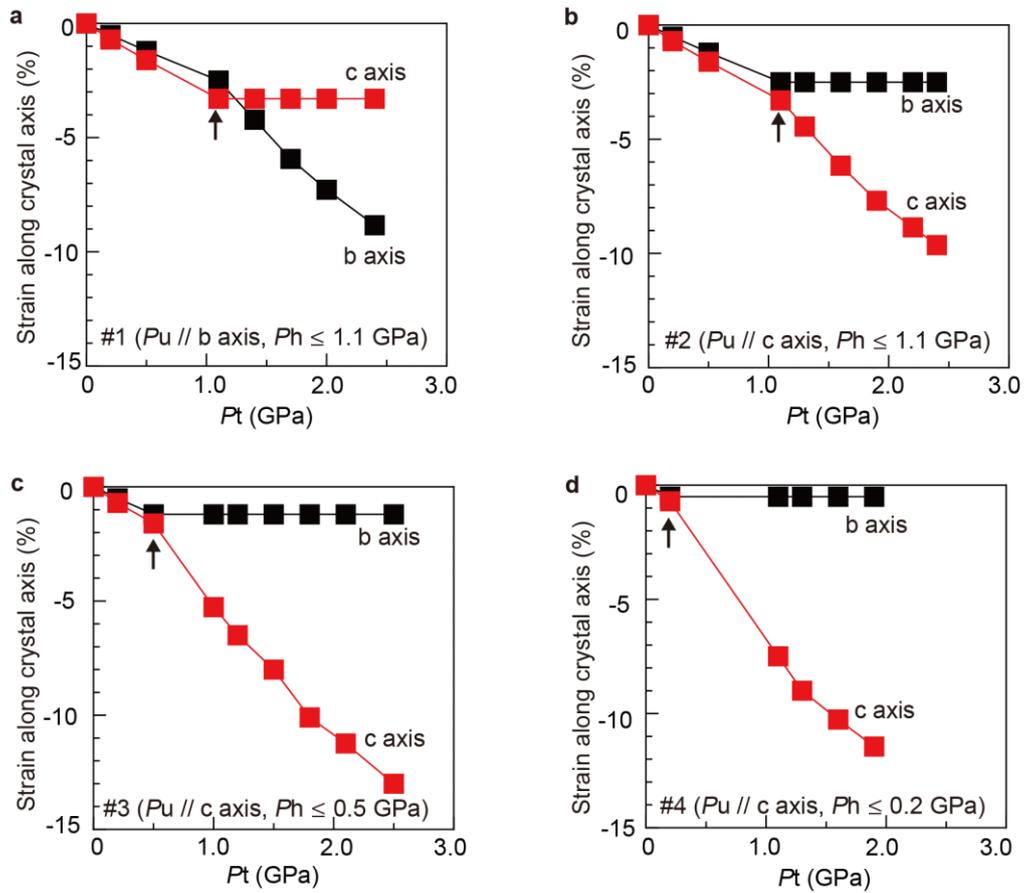

**Supplementary Fig. 4 | Pressure dependence of estimated strain values along b and c axis for sequence #1 (a), #2 (b), #3 (c), and #4 (d).** The applied pressure is hydrostatic below the pressure pointed by arrows, and uniaxial above them.

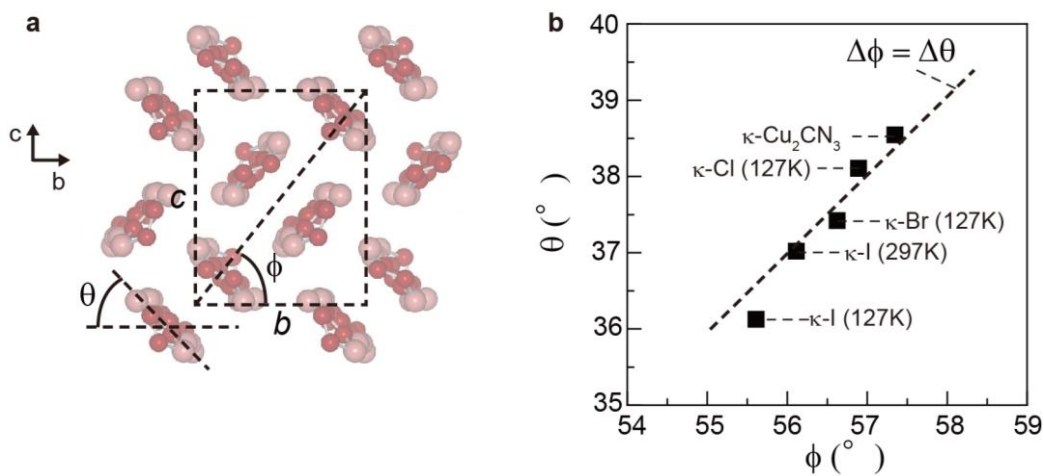

**Supplementary Fig. 5 | Relationship between the orientation of molecules and the ratio of lattice parameters. a,** Molecular arrangement in b-c plane. $\phi$ and $\theta$ are arctan(*c*/*b*) and the angle between short axis of the molecule and b axis. **b,** $\phi$ vs $\theta$ in κ-ET compounds, κ-(ET)$_2$Cu$_2$(CN)$_3$ (κ-Cu$_2$CN$_3$), κ-(ET)$_2$Cu[N(CN)$_2$]Cl (κ-Cl), κ-(ET)$_2$Cu [N(CN)$_2$]Br (κ-Br), and κ-(ET)$_2$Cu[N(CN)$_2$]I (κ-I).

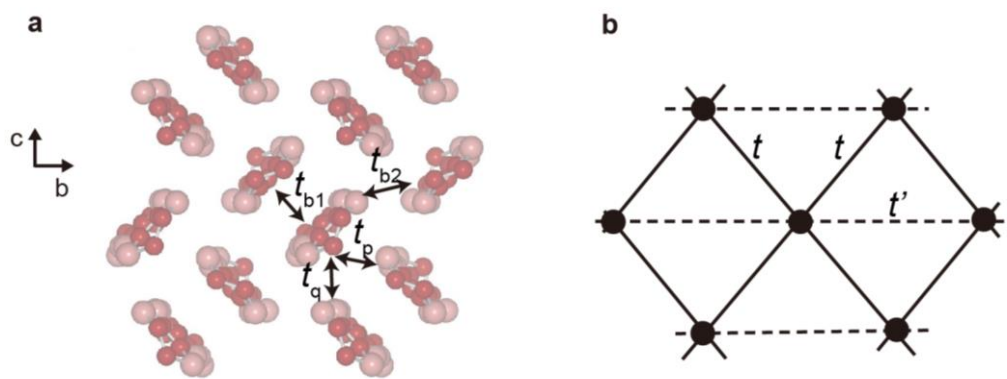

**Supplementary Fig. 6 | Transfer integrals between molecules (a) and dimers (b).**

The inter-dimer transfer integrals $t$ and $t'$ are expressed as $t = (-t_p + t_q)/2$ and $t' = -t_{b2}/2$ using the inter-molecular transfer integrals $t_p$, $t_q$, and $t_{b2}$.

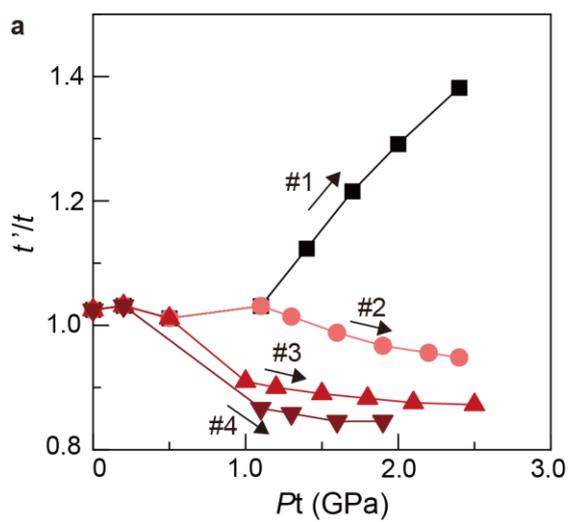 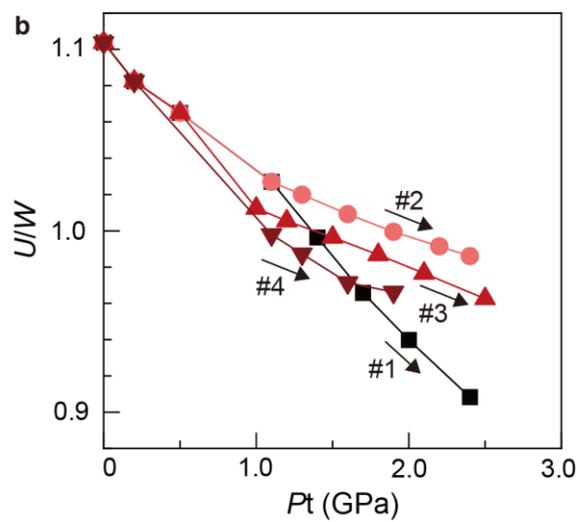

**Supplementary Fig. 7 | Estimated band parameters for sequences #1-4. a,** Pressure dependence of *U*/*W*. **b,** Pressure dependence of *t'*/*t*.